\documentclass[conference]{IEEEtran}
\IEEEoverridecommandlockouts

\usepackage{amsmath,amssymb,amsfonts}
\usepackage[hidelinks]{hyperref}
\usepackage{graphicx}
\usepackage{subfig}
\usepackage{xcolor}
\usepackage[ruled,vlined]{algorithm2e}
\usepackage{listings}
\begin{document}

\title{A Survey of Singular Value Decomposition Methods for Distributed 
Tall/Skinny Data}

\author{\IEEEauthorblockN{Drew Schmidt}
\IEEEauthorblockA{\textit{Oak Ridge National Laboratory} \\
Oak Ridge, Tennessee\\
schmidtda@ornl.gov}
}

\maketitle

\begin{abstract}
The Singular Value Decomposition (SVD) is one of the most important matrix 
factorizations, enjoying a wide variety of applications across numerous 
application domains. In statistics and data analysis, the common applications of 
SVD such as Principal Components Analysis (PCA) and linear regression. Usually 
these applications arise on data that has far more rows than columns, so-called
``tall/skinny'' matrices. In the big data analytics context, this may take the 
form of hundreds of millions to billions of rows with only a few hundred 
columns. There is a need, therefore, for fast, accurate, and scalable 
tall/skinny SVD implementations which can fully utilize modern computing 
resources. To that end, we present a survey of three different algorithms for 
computing the SVD for these kinds of tall/skinny data layouts using MPI for 
communication. We contextualize these with common big data analytics 
techniques, principally PCA. Finally, we present both CPU and GPU timing 
results from the Summit supercomputer, and discuss possible alternative 
approaches.
\end{abstract}

\section{Introduction}
\label{sec:introduction}

To call the Singular Value Decomposition (SVD) useful is an extraordinary 
understatement. It is difficult to think of another matrix factorization that 
has enjoyed more applications of more importance than the SVD. Perhaps this is 
because many of the things one may wish to do numerically with matrices can be 
done via SVD. To name a few of the more important ones, one can compute matrix 
inverses, solve a system of equations, calculate determinants, compute least 
squares solutions to overdetermined systems, compute condition numbers, norms, 
column rank, and on and on. It should be no surprise then that many of the 
applications of matrices to real world problems are either solved outright by 
SVD or are simpler to solve because of it. 

Data analytics and statistics are not immune from the dominance of this 
important factorization. One useful application of SVD is in fitting linear and 
generalized linear models (LM and GLM). The standard ``linear regression'' 
formulation is essentially just linear least squares, so this is often solved by 
a QR factorization alone. However, there are some advantages to using SVD 
instead. Indeed, SVD is a rank-revealing factorization, and some popular QR 
algorithms are not. This is important for statisticians because in the context 
of statistical experimental design, their so-called model matrices are 
sometimes rank-degenerate by construction. As for GLM, a good LM fitter can be 
used to fit a GLM using iteratively re-weighted least 
squares~\cite{mccullagh1989generalized}.

Another popular application of SVD in the statistical and data sciences is 
Principal Components Analysis or PCA~\cite{wall2003singular}. PCA is an 
exploratory data analysis tool which is often used to visualize a 
high-dimensional dataset in two or three dimensions while maximizing the amount 
of ``information'' (in this case, variability) retained. The dataset consists of 
a matrix whose columns are variables and whose rows are observations. In this 
setting, the data is first mean-centered column-by-column, and then projected 
onto the right singular vectors (alternatively, the right singular vectors are 
``removed''). This rotation orders the newly constructed variables in 
decreasing order of variability retained from the original dataset. So plotting 
only the first few gives a good tradeoff between preserving all of the 
variability of the original dataset and not being able to visualize it at all.

For data analytics and statistics, dataset are typically of the tall/skinny 
variety. It is not uncommon, particularly in big data applications, to have 
only a few hundred or a few thousand columns, but hundreds of millions to 
billions of rows. In fact, while tall/not-so-skinny applications do arise, 
square or nearly square matrices are essentially unheard of. Some domains such 
as bioinformatics have the transposed problem of short/wide matrix shapes. Much 
of the mathematics of tall/skinny data works out with appropriate transpose 
arithmetic and symbol-pushing. So for the sake of brevity, we only focus on the 
tall/skinny case.

In the sections that follow, we will present several algorithms for computing 
SVD. The algorithms we describe are not new to this article. However, we 
provide some implementation details which are often left as an exercise to the 
reader. In particular, we stress the issues that arise in big data contexts, 
where data is distributed across multiple processes. Our implementations use 
MPI~\cite{MPI1994} for managing communication. Of course, much of the big 
data analytics world has focused on using the MapReduce 
algorithm~\cite{dean2010mapreduce} instead. For example, 
\cite{constantine2011tall} takes an in-depth look at developing a tall/skinny QR 
using MapReduce, which in many ways is very similar to what we present in 
Section~\ref{subsec:qr}.

So why bother with MPI? For as often maligned as it is, we argue that it has 
several serious advantages over MapReduce, even for big data analytics. One 
disadvantage of the MapReduce approach is that it uses languages and programming 
models alien to many in the statistical and mathematical sciences. For example, 
forcing all computation through the key/value pair mapper/reducer model is 
usually cumbersome and unnatural for matrix computations. One may challenge this 
as a matter of taste. However, we would point out that no one programs in this 
way among the large variety of programming languages and analysis packages that 
implement matrix calculations (Matlab, R, NumPy, Julia, \dots). One programs in 
this way with MapReduce only because they must. In contrast, the Single 
Program/Multiple Data or SPMD model~\cite{darema2001spmd} common to MPI 
programs is a natural extension of serial programming. It is fair to say that 
MapReduce makes parallelism easy, in that it is never explicit. But this comes 
at the cost of making everything else harder.

Perhaps more importantly, MPI allows us to implement our matrix computations of 
interest with significantly less communication overhead \emph{inherently}. 
Indeed, each iteration of MapReduce includes a shuffle operation that is 
equivalent to an \texttt{MPI\_Alltoall} call~\cite{tu2008scalable, 
plimpton2011mapreduce}. And any given implementation may require multiple 
iterations, meaning multiple shuffles. However, for our SVD computations, we are 
generally able to use a single \texttt{MPI\_Allreduce} call.

Finally, most MapReduce users today use the Apache Spark 
framework~\cite{zaharia2016apache}. This is a very heavy dependency which 
can have problems integrating with shared resources like clusters and 
supercomputers. And the performance of Spark is known to be poor compared to 
high performance computing solutions~\cite{xenopoulos2016big}.

For all of these reasons, we will only proceed with an MPI-based communication 
strategies. But before we can properly begin, we outline some background 
information.
\section{Background}
\label{sec:background}

\subsection{Basic Definitions}
Let $A$ be a real-valued matrix with $m$ rows and $n$ columns. Then we take the 
SVD of $A$ to be:
\begin{align*}
A_{m\times n} = U_{m\times r} \Sigma_{r\times r} V_{n\times r}^T
\end{align*}
where $r$ is usually the minimum of $m$ and $n$, although it may be taken to be 
the rank of $A$ which is no greater than the minimum of $m$ and $n$. This is 
sometimes called the "compact SVD". Additionally, $\Sigma$ is a diagonal matrix, 
and $U$ and $V$ are orthogonal. To say that a matrix is orthogonal means that 
its transpose is its inverse. For rectangular matrices (as is often the case 
for SVD), when we say that a matrix is orthogonal, we only mean so along the 
short dimension. So if $U$ is as above with $m>n\geq r$, then $U^TU = 
I_{r\times r}$. However, $UU^T \neq I_{m\times m}$ since the rows of $U$ can not 
be linearly independent.

For the remainder, we will always assume that $m>n$.
\subsection{Truncated SVD}
Sometimes only a few singular values/vectors are required for a given 
application, as is often the case with PCA. However, because of limitations 
inherent to the algorithms that compute them: if you want one, you get them all. 
This can add significant runtime and memory overhead to calculate data that 
will simply be thrown away.

Although it is nearly 100 years old, the so-called power method is still used 
today for truncated problems. We do not include the details of the method here, 
but there are many good surveys on the topic, for 
example~\cite{booth2006power}. Another common technique for truncated SVD is the 
Lanczos method~\cite{lanczos1950iteration}, and its many variants. Both of 
these classes of methods are iterative solvers which are in actual fact 
techniques for spectral decomposition, not SVD. That is, these methods provide 
approximations for the largest eigenvalues and their corresponding eigenvectors. 
There is a close relationship between SVD and the eigenproblem via the so-called 
``normal equations matrix,'' which we discuss in depth in 
Section~\ref{subsec:normaleqns}. But there is another relationship that is 
exploited specifically for Lanczos methods. For a rectangular matrix $A$ of 
dimension $m\times n$ with $m>n$, you must first ``square up'' the matrix:
\begin{align*}
H := 
\begin{bmatrix}
  0 & A \\\\
  A^T & 0
\end{bmatrix}
\end{align*}
Then $H$ is a square, symmetric matrix of order $m+n$, so you can perform the 
Lanczos method $k$ times to approximate the eivenvalues of $H$. Note that for 
full convergence, we need to run the iteration $m+n$ times, not $n$ (which would 
be a very expensive way of computing them). The approximations to the singular 
values of $A$ are given by the non-zero eigenvalues of $H$, and the singular 
vectors (left and right) are found in $Y_k := \sqrt{2}\thinspace Q_k S_k$ 
(Theorem 3.3.4 of~\cite{demmel1997applied}). Specifically, the first $m$ rows of 
$Y_k$ are the left singular vectors, and the remaining $n$ rows are the right 
singular vectors.

For the purposes of computing SVD, the Lanczos method primarily relies 
on a series of matrix-vector products involving $H$. Because of this and 
the block anti-diagonal structure of $H$, one does not ever need to explicitly 
form it in computer memory. The reliance of this technique on nothing more 
complicated than matrix-vector products becomes particularly valuable if $A$ 
is itself sparse.

One common application of the Lanczos method for data science is Latent 
Semantic Indexing~\cite{berry1995using}, which comes from text analysis. It 
involves computing the truncated SVD of a generally very sparse matrix called 
the term frequency-inverse document frequency (tf-idf) matrix. The Lanczos 
method and the power iteration are sometimes used in implementations of the 
PageRank algorithm~\cite{page1999pagerank}, which is a technique to find 
influential nodes in a graph. Although this example is about eigendecomposition, 
not strictly SVD. It is worth noting, however, is that these are both sparse 
matrix problems.

More recently, there is the method of computing truncated SVD via random
projections~\cite{halko2011finding}. This technique works well for dense and 
can be easily implemented for distributed data problems, and as such is the 
subject of Section~\ref{subsec:rsvd}.
\subsection{Computing Assumptions}
\label{subsec:bgcompute}
In the following section, our implementation emphasis is on distributed data. We 
use the term ``distributed'' and ``distribution'' in the computing sense, not 
the statistical sense. Specifically, the data is a matrix split across processes 
in a one-dimensional (1-d) fashion, so that processes in the MPI communicator 
own contiguous blocks of rows. If a process owns part of a row, it owns the 
entire row. If we have $p$ processes in the MPI communicator, we conceptually 
split a matrix $A$ as:
\begin{align}
\label{math:1d}
A = 
\begin{bmatrix}
 A_1 \\
 A_2 \\
 \vdots \\
 A_p
\end{bmatrix}
\end{align}
where $A_1$ is the block of rows on MPI rank 0, $A_2$ on rank 1, and so on.

We assume that any time a matrix is ``large'', it is distributed, and otherwise 
it is common to all processes. So for example, if $A$ is a distributed matrix 
with many rows and few columns and if we have $A = U\Sigma V^T$, then $U$ 
is distributed in identical fashion to $A$, but $\Sigma$ and $V$ are both small 
and so every process owns a copy of each.

Matrix multiplications always take the form of a distributed matrix times a 
non-distributed matrix resulting in a distributed matrix, or the transpose of a 
distributed matrix times a distributed matrix, resulting in a non-distributed 
matrix. The former is an entirely local computation, and the latter is a local 
matrix product followed by an \texttt{MPI\_Allreduce}.

We have noted that communication will be handled by MPI. As for the local 
computations, we rely on high-quality implementations of the de facto 
standards, the BLAS~\cite{lawson1979basic} and LAPACK~\cite{lug}. For example, 
in our benchmarks in Section~\ref{sec:benchmarks}, we use 
OpenBLAS~\cite{xianyi2012openblas} for computations on CPU. For GPUs, we use the 
CUDA environment~\cite{kirk2007nvidia}, including their BLAS and LAPACK 
semi-clones cuBLAS and cuSOLVER, although we discuss alternate 
strategies in Section~\ref{sec:conclusions}.

When referring to some specific local computations which are to be performed by 
an LAPACK implementation, we will refer to the function name without the type 
character. So for example, when referring to the symmetric eigensolver, we use 
\texttt{syevr}, with the understanding that the ``s'' or ``d'' variant will be 
used appropriately. Likewise for a QR, \texttt{geqrf}, and for a local SVD, 
\texttt{gesdd}. We will also make regular use of the shorthand \texttt{qr\_Q} 
and \texttt{qr\_R} to refer to functions which compute the $Q$ or $R$ matrices 
of a QR factorization, respectively. In algorithmic descriptions, the 
implementation details of these functions will be noted parenthetically if at 
all ambiguous. A local matrix product is always assumed to be computed by a BLAS 
\texttt{gemm} operation. 
\section{Algorithms}
\label{sec:algos}

\subsection{Method 1: The Normal Equations}
\label{subsec:normaleqns}

Given an over-determined system of equations $Ax=b$, the ``normal equation'' is
\begin{align*}
A^TAx = A^Tb 
\end{align*}
For this reason, $A^TA$ is sometimes referred to as ``the normal equations 
matrix'', or even sometimes quite incorrectly as ``the normal equations.'' In 
statistics, this is sometimes called the ``crossproducts matrix''. In fact, in 
the popular statistics and data programming language R~\cite{team2000r}, this 
operation is performed by the \texttt{crossprod} function.

Because $A^TA$ is symmetric, we know that it must have an eigendecomposition $A 
= V \Lambda V^T$ where $\Lambda$ is the diagonal matrix of eigenvalues and $V$ 
is an orthogonal matrix of eigenvectors. But taking the SVD of $A$ we have
\begin{align*}
A^TA &= \left( U \Sigma V^T \right)^T \left( U \Sigma V^T \right) \\
  &= V \Sigma U^T U \Sigma V^T \\
  &= V \Sigma^2 V^T
\end{align*}
So the eigenvalues of $A^TA$ are the square of the singular values of $A$, and 
the right singular vectors of $A$ form a set of eigenvectors of $A^TA$. If 
we know $\Sigma$ and $V$ then we can recover $U$ easily since $U = A V 
\Sigma^{-1}$.

If $A$ is distributed as in Formula~\eqref{math:1d}, then each MPI rank $i$ 
can compute the local $N_i = A_i^T A_i$. Then we merely need add up all 
the entries of all the local $N_i$ matrices across all the ranks using a call 
to \texttt{MPI\_allreduce}. We summarize this procedure in 
Algorithm~\ref{algo:cpsvd}.

\begin{algorithm}[t]
\SetAlgoLined
\KwData{1-d distributed real matrix $A$ with $m>n$}
\KwResult{$\Sigma$, optionally $U$ and $V$}
Compute $N_{local} = A_{local}^T A_{local}$\;
Compute $N$ = \texttt{MPI\_allreduce}($N_{local}$)\;
Compute the eigenvalues $\Lambda$ (optionally the eigenvectors $V$) of $N$ 
locally via \texttt{syevr}\;
Let $\sigma_i = \sqrt{\lambda_i}$ for $i=1, \dots, n$\;
\If{compute $U$}{
  $U = AV\Sigma^{-1}$\;
}
\caption{SVD Via Normal Equations Matrix}
\label{algo:cpsvd}
\end{algorithm}

This approach has several advantages for the tall/skinny case. For one, it is 
extremely simple to implement. Additionally, because it is dominated by a 
matrix-matrix multiplication, it is extremely fast. The downside is that it may 
have numerical issues. Indeed, explicitly forming the normal equations matrix 
squares the condition number of $A$. For applications like linear regression, 
this may be a serious problem. For PCA, it probably does not matter.

If the data resides on a GPU and the MPI implementation supports GPUDirect, 
then the \texttt{MPI\_Allreduce} computation can be done as-is. Otherwise, a 
CPU buffer will be required.
\subsection{Method 2: QR}
\label{subsec:qr}

We can factor
\begin{align*}
A_{m\times n} = Q_{m\times n} R_{n\times n}
\end{align*}
where $R$ is upper triangular, and $Q$ is orthogonal. Strictly speaking, this 
is the ``thin'' QR factorization of $A$~\cite{golub1996matrix}. However, we 
will never need or make additional reference to the full factorization.

With $A=QR$, if we next take the SVD of $R$ then we have
\begin{align*}
A &= QR \\
  &= Q \left( U_R \Sigma_R V_R^T \right) \\
  &= \left( Q U_R \right) \Sigma_R V_R^T
\end{align*}
Then $U\Sigma V^T$ is a singular value decomposition of $A$, where $U = Q U_R$ 
(which is orthogonal, being a product of orthogonal matrices), $\Sigma = 
\Sigma_R$ and $V = V_R$. So if we can compute the right factor $R$ of $A$, then 
we can compute its singular values and right singular vectors. As before, we 
can recover the left singular vectors if desired from the identity $U = A V 
\Sigma^{-1}$.

The key idea behind computing $R$ is that given a matrix split by rows as in 
Formula~\eqref{math:1d}, computing $R$ of $A$ can be achieved by computing the 
$R$ matrix of the pieces, stacking them two at a time, and then computing their 
$R$ matrix. This process continues until all of the pieces have contributed. 
This is known as a communication-avoiding QR, or 
CAQR~\cite{demmel2008communication, agullo2010qr}, and it is a particular kind 
of tall/skinny QR, or TSQR.

More formally, if we split the rows of $A$ into $A_1$ and $A_2$ so that $A = 
\begin{bmatrix}A_1 \\ A_2 \end{bmatrix}$, then we can factor $A_i = Q_i R_i$ 
($i=1, 2$). Then
\begin{align*}
A &= 
  \begin{bmatrix} Q_1 & 0 \\ 0 & Q_2 \end{bmatrix}
  \begin{bmatrix} R_1 \\ R_2 \end{bmatrix}
\end{align*}
This is not yet a QR factorization of $A$, because the right factor is not 
upper triangular. Denote the left factor as $\tilde{Q}$ and let $\hat{Q}R$ be 
a QR factorization of the right factor. Then $A = \tilde{Q} \hat{Q} R$. And 
because $\tilde{Q}$ and $\hat{Q}$ are orthogonal, so is their product. And 
hence this is a QR factorization of $A$.

We can cast this as a binary operation which accepts two $R$ matrices, stacks 
them, computes and returns their $R$ matrix. This binary operation is always 
associative, and under some circumstances it may be 
commutative~\cite{agullo2010qr}. This allows us to create a custom MPI reduce 
operation, which can take advantage of MPI's optimizations like recursive 
doubling to minimize communication~\cite{rabenseifner2004optimization}. We 
summarize this in Algorithm~\ref{algo:qr_reducer}.

\begin{algorithm}[t]
\SetAlgoLined
\KwData{Real matrices of order $n$ $R_{in}$ and $R_{out}$}
Let $T := \begin{bmatrix} R_{in} \\ R_{out} \end{bmatrix}$\;
Set $R_{out} = $ \texttt{qr\_R}($T$)\;
\caption{QR Reducer}
\label{algo:qr_reducer}
\end{algorithm}

\begin{algorithm}[t]
\SetAlgoLined
\KwData{1-d distributed real matrix $A$ with $m>n$}
\KwResult{$\Sigma$, optionally $U$ and $V$}
Compute $R_{local} = $ \texttt{qr\_R}$(A_{local})$ via \texttt{geqrf}\;
Compute $R$ = \texttt{QR\_allreduce}($R_{local}$)\;
Compute $\Sigma$ and optionally $V$ via \texttt{gesdd}. Recover 
$U=AV\Sigma^{-1}$ if desired.\;
\caption{SVD via CAQR}
\label{algo:tsqr}
\end{algorithm}

For floats, a C-like implementation may look like:
\begin{lstlisting}[language=C++]
void qr_reducer(void *Rin, void *Rout, int *len, MPI_Datatype *dtype);

MPI_Datatype qrtype;
MPI_Type_contiguous(n*n, MPI_FLOAT, qrtype);
MPI_Type_commit(qrtype);

MPI_Op qrop;
MPI_Op_create((MPI_User_function*) qr_reducer, 1, &qrop);
\end{lstlisting}
The reducer can then be called as follows:
\begin{lstlisting}[language=C++]
MPI_Allreduce(Rin, Rout, 1, qrtype, qrop, comm);
\end{lstlisting}

We summarize the use of the QR reducer to implement the CAQR in 
Algorithm~\ref{algo:tsqr}. As a final note, there are some subtleties here if 
the data is a GPU device pointer. Specifically, at this time on the compute 
platform we run our experiments on, GPUDirect does not support custom 
operations. So we must use CPU buffers for transferring the data. This creates 
some unavoidable overhead in copying the two input $R$ matrices from host to 
device, and then the emitted $R$ matrix from device to host.
\subsection{Method 3: Truncated SVD via Random Projections}
\label{subsec:rsvd}

Finally, we present a truncated SVD algorithm from the famous ``Finding 
Structure with Randomness''~\cite{halko2011finding}. The details are more 
involved than the above, so we do not repeat them here. We note that the key 
ideas come from the Prototype for Randomized SVD and and Algorithm~4.4 pieces 
of the paper.

Since this is a truncated SVD, we specify an additional parameter $k$ 
indicating the number of singular values/vectors to compute. There is also a 
parameter $q$ which is an exponent, but which we treat as a hyper-parameter. 
Finally, there is a random matrix $\Omega$ used internally to the algorithm 
for initialization (the random projection). The original authors use data 
generated from a standard normal distribution. But we find experimentally that 
random uniform works just as well, and with the advantage of being slightly 
faster to generate.

The computations involved are a series of matrix products and computing $Q$ 
matrices from a QR factorization. Some of the matrix products are 
communication-free (see the discussion in Section~\ref{subsec:bgcompute} for 
details), and some of the $Q$ computations are also purely local. For the $Q$ 
computations which operate on distributed data, a TSQR should be used. 
Naturally, we can use the CAQR from Algorithm~\ref{algo:tsqr} of 
Section~\ref{subsec:qr}. We summarize the details in Algorithm~\ref{algo:rsvd}.

\begin{algorithm}[t]
\SetAlgoLined
\KwData{1-d distributed real matrix $A$ with $m>n$, integers $k<n$ and $q$}
\KwResult{$\Sigma$, optionally $U$ and $V$}
Generate $\Omega_{n\times 2k}$ random uniform or standard normal\;
Let $Y_{m\times 2k} = A\Omega$\;
Compute $Q_Y = $ \texttt{qr\_Q}$(Y)$ \space\space (CAQR)\;
\For{$i\gets 0 $ \KwTo $q$}{
  $Z_{n\times 2k} = A^T Q_Y$\;
  $Q_Z = $ \texttt{qr\_Q}$(Z)$ \space\space (local)\;
  $Y = A Q_Z$\;
  $Q_Y = $ \texttt{qr\_Q}$(Y)$ \space\space (CAQR)\;
}
Let $B_{2k\times n} = Q_Y^T A$\;
Local SVD of $B$ gives approximation for $\Sigma$, and optionally $V^T$. 
Recover $U = Q_Y U_B$ if desired.
\caption{Randomized SVD}
\label{algo:rsvd}
\end{algorithm}

Finally, this technique has been successfully employed for truncated PCA on 
large datasets~\cite{schmidt2017programming} using ScaLAPACK~\cite{slug}. We 
will compare our TSQR via CAQR approach to one using ScaLAPACK for the 
intermediate calculations in Section~\ref{sec:benchmarks}.
\section{Experiments and Discussion}
\label{sec:benchmarks}

Below we present the results of several experiments. We run all of our 
benchmarks on Summit, which at the time of writing is ranked second on the Top 
500 list~\cite{dongarra1997top500,top500june2020}. Summit is an IBM system 
managed by the Oak Ridge Leadership Computing Facility at Oak Ridge National 
Laboratory. It has 4608 nodes, each equipped with two 22-core IBM POWER9 CPUs, 
and six NVIDIA Tesla V100 GPUs. The GPUs are connected by NVIDIA's high-speed 
NVLink interconnect, and nodes are connected by a high-performance Mellanox 
InfiniBand network. The CUDA-Aware MPI from IBM (Spectrum MPI) allows for some 
GPUDirect communication, which we exploit whenever possible.

For the benchmarks, we present both CPU and GPU implementations of the three 
algorithms presented in Section~\ref{sec:algos}. We will refer to them as cpsvd 
for crossproducts-svd, tssvd for tall/skinny CAQR-based SVD, and rsvd for 
randomized SVD, being implementations of the algorithms from 
Sections~\ref{subsec:normaleqns}, \ref{subsec:qr}, and \ref{subsec:rsvd}, 
respectively. For each, we measure only the time to compute the 
singular values (or their approximations), since that is the only portion 
requiring communication. We would expect some separation of the CPU and GPU 
timings in favor of GPUs if additionally computing the singular vectors.

As we evaluate the various implementations throughout, we vary the sizes of the 
inputs, and will describe them in detail as we describe each particular result. 
In each case, we use random data from a standard normal distribution. We also 
present results for both 32-bit and 64-bit floating point numbers (C++ 
\texttt{float} and \texttt{double}). In each case, we operate on the same size 
matrix in bytes, using half the number of rows for the 64-bit data as the 32-bit 
data. All matrices are chosen to have 250 columns. This is somewhat arbitrary, 
but it gives a tall/skinny matrix, and in data analysis, the number of columns 
is often on the order of numbering no more than a few hundred. Another constant 
throughout all experiments is that any time we present a result for the 
randomized SVD, we use $k=2$, $q=2$, and we use random uniform data for the 
projection.

\begin{figure}[t]
  \includegraphics[width=.475\textwidth]{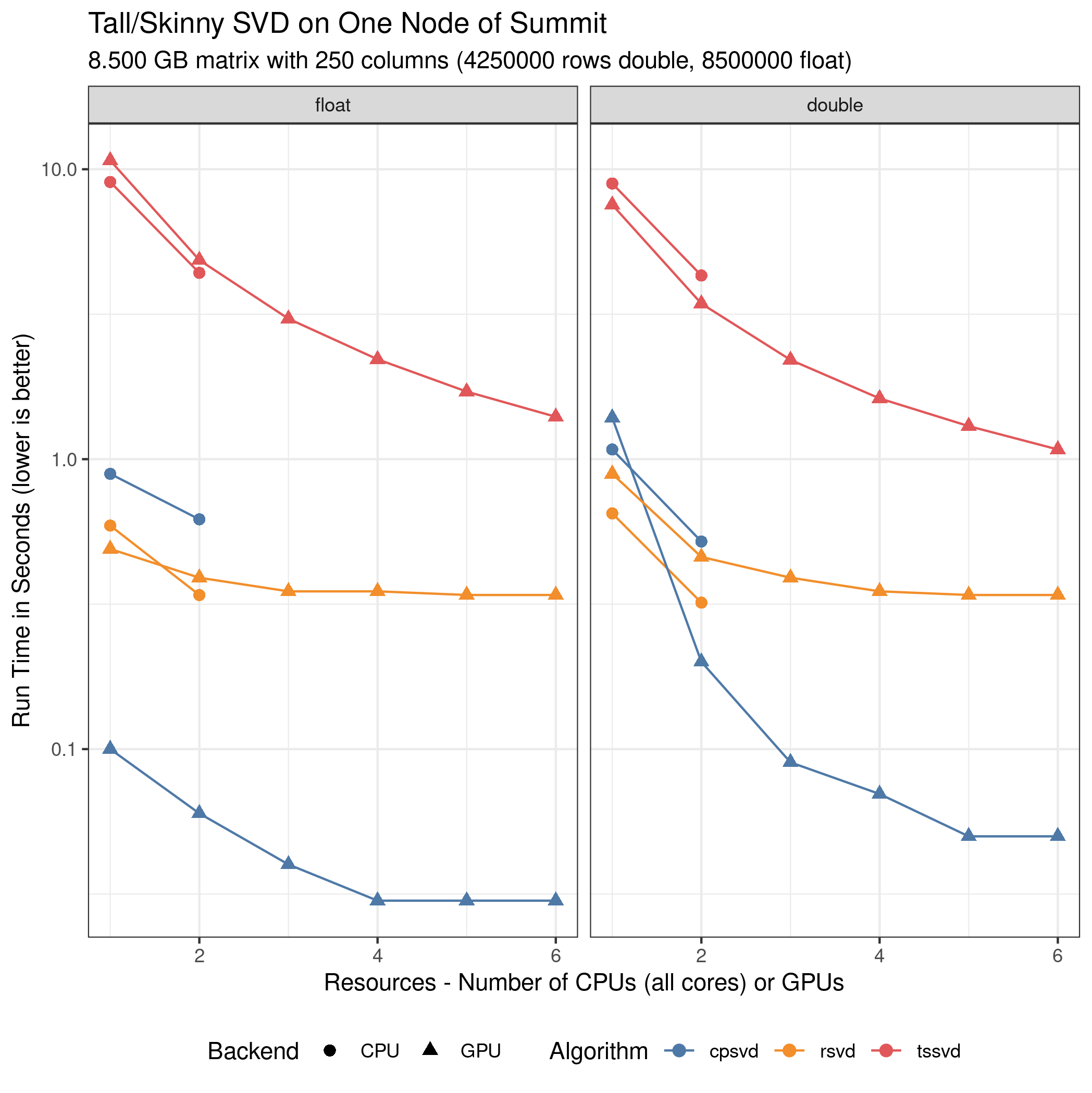}
  \caption{SVD benchmarks on one Summit node. The x-axis shows the number of 
resources; for CPU this is all cores (no hyperthreads) on each physical CPU. 
Summit has two physical CPUs and six GPUs per node.}
  \label{fig:1node}
\end{figure}

We begin by examining the performance of the of three implementations on a 
single node, with results presented in Figure~\ref{fig:1node}. This is a strong 
scaling benchmark, where we fix a data size and increase the number of 
resources. Each Summit node has two physical CPUs and six GPUs, hence the 
disparity in the number of timing results for each.

Let us begin by examining the float vs double performance. Recall that the data 
size for each test is the same (twice as many rows for float as double), so we 
would expect the two to be roughly the same. And indeed, this holds true for 
all CPU timings. However, there are some interesting differences for GPU. The 
rsvd on one GPU completes in roughly half the time for float than for 
double. We are unable to dismiss this as mere sampling variation, as re-running 
the experiment multiple times on different nodes produces similar results. 

However, the cpsvd results may shed some light on what is occurring. Notice 
here the large discrepancy between the GPU run times across the two fundamental 
types. Examination of the benchmark by the NVIDIA Nsight 
profiler~\cite{nsight20133} shows that 95\% of the total run time is dominated 
by the random generator prior to launching the cpsvd kernel together with a 
matrix multiplication, carried out by cuBLAS. The striking difference in 
performance suggests that perhaps the \texttt{sgemm} variant is using the 
tensor cores of the GPU. We evaluated the benchmark with various program level 
and environment level settings, but were unable to affect a different result in 
the timings. We contacted an NVIDIA support engineer with deep familiarity with 
Summit for clarification; but if the root of the issue was discovered, it was 
not shared with the author.

The relative consistency for the tssvd, which has very few matrix 
multiplications, compared to the rsvd which has several, then finally compared 
to the cpsvd which is essentially only a matrix multiplication, it seems 
reasonable to conclude that the performance difference we see here is due to 
acceleration by the tensor cores. Why we are unable to disable this is still 
unclear, however. For a detailed list of our software environment, see 
Section~\ref{sec:artifacts}.

Next, let us compare the CPU vs GPU performance. The discussion above colors 
the evaluation of cpsvd and rsvd. As for tssvd, the run times are fairly close 
to each other for CPU and GPU. This is likely because of the additional 
overhead required moving data back and forth between host and device memory 
during the \texttt{QR\_allreduce} computation, as discussed at the end of 
Section~\ref{subsec:qr}. Also, the local problem at each step of the reduction 
operation amounts to stacking the two $n\times n$ matrices on top of each other, 
then emitting the $R$ matrix computed from the QR factorization of the $2n\times 
n$ matrix. With our fixed $n=250$, this problem size is fairly small, so we 
never get to take full advantage of the GPU flops. Specifically, each local QR 
operates on only 1 MB in double precision and 0.5 MB in single. This likely 
explains why the double precision GPU version performs slightly better than its 
single precision variant. One final thing worth noting is that the CPU cores on 
Summit are extremely fast, making them much more competitive in general.

\begin{figure}[t]
  \includegraphics[width=.475\textwidth]{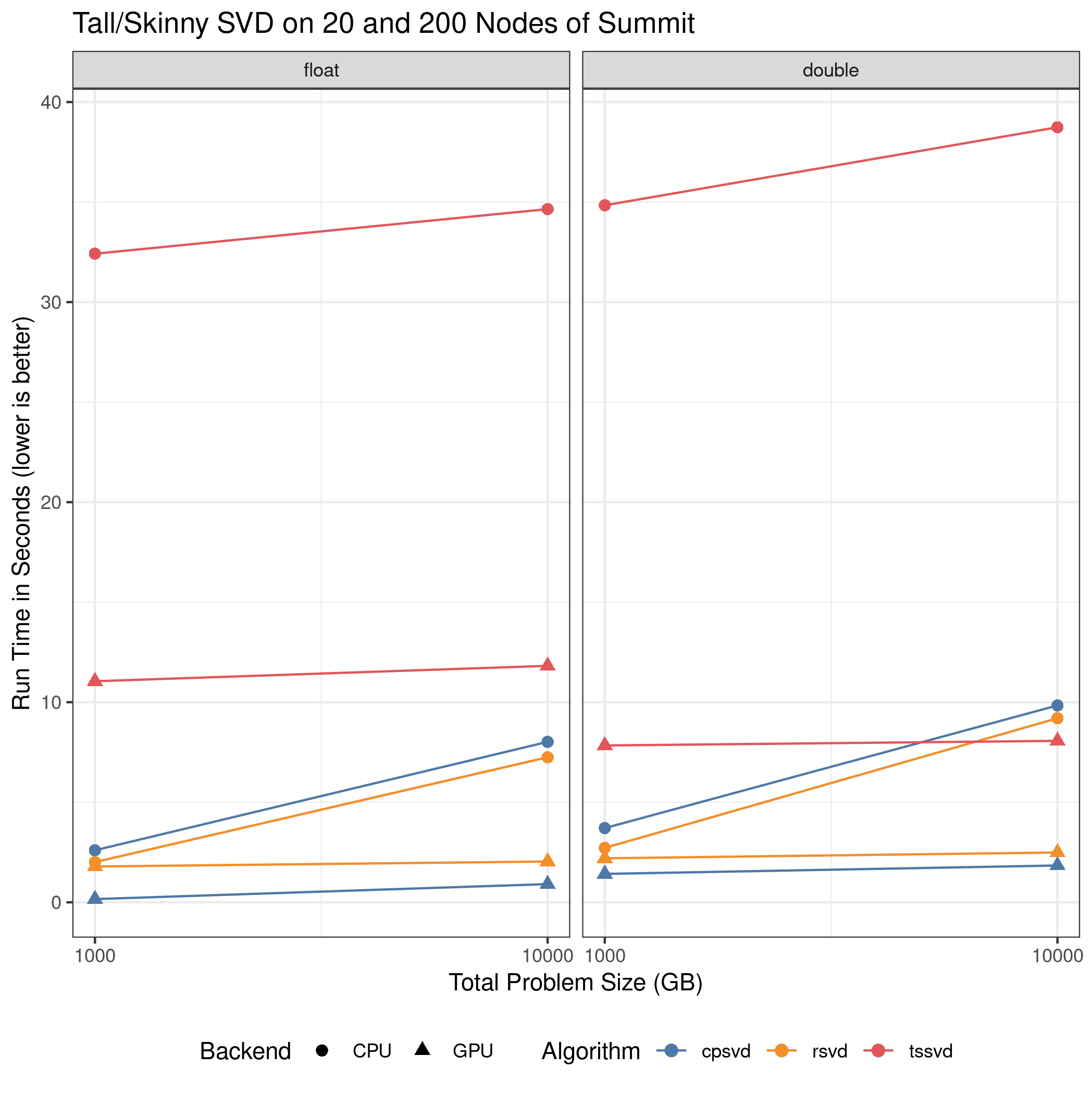}
  \caption{SVD benchmarks on 20 and 200 Summit nodes, on problems of size 1 TB 
and 10 TB, respectively. Results here are node-level performance. For example, 
20 nodes with a CPU backend is 40 total physical CPUs (all cores, no 
hyperthreads), and at 20 nodes with a GPU backend is 120 total GPUs.}
  \label{fig:weak}
\end{figure}

Next we examine how each of the implementations scales in the weak sense. The 
results are presented in Figure~\ref{fig:weak}, and they show the timings from 
1 TB and 10 TB total problems at 20 and 200 nodes, respectively. This works out 
to a per-GPU local problem size close to that in our first experiment (here 
8.333 GB vs 8.5 GB formerly). At 20 nodes, each matrix has $5\cdot 10^8$ rows 
for double precision data and $10^9$ for single, and at 200 nodes each matrix 
has $5\cdot 10^9$ rows for double precision data and  $10^{11}$ rows for single. 
One notable observation before proceeding is that for the smaller data size, 
each index is 32-bit addressable, while the number of rows is not 32-bit 
addressable for the larger matrices. This is an issue we will return to in the 
next experiment.

Proceeding as before, if we first examine the plot for a comparison of float vs 
double performance, we find that the analysis above appears to still hold. So 
for the sake of brevity we do not repeat it here. Comparing CPU vs GPU 
performance, we first note that this is a node-level comparison, compared to 
the above which was resource-level. For the CPU times, this accounts for 40 
CPUs for the smaller size and 120 at for the larger one. Likewise, this is 400 
GPUs at for the smaller size and 1200 for the larger. And indeed, what we see 
is a roughly three-fold performance difference over what we saw above in 
Figure~\ref{fig:1node}. The only observation of note is that the GPU run times 
are all largely flat, which is ideal in this case, while the CPU variants have 
worse scaling. We do not immediately understand why this is so. It is possible 
that because Summit is essentially a GPU machine, the vendor MPI library may 
not be well-optimized for intra-node CPU communication. The total number of MPI 
ranks is 840 and 8400 for CPU, but only 120 and 1200 for GPU. Again we note 
that a full list of our software environment is provided in 
Section~\ref{sec:artifacts}.

\begin{figure}[t]
  \includegraphics[width=.475\textwidth]{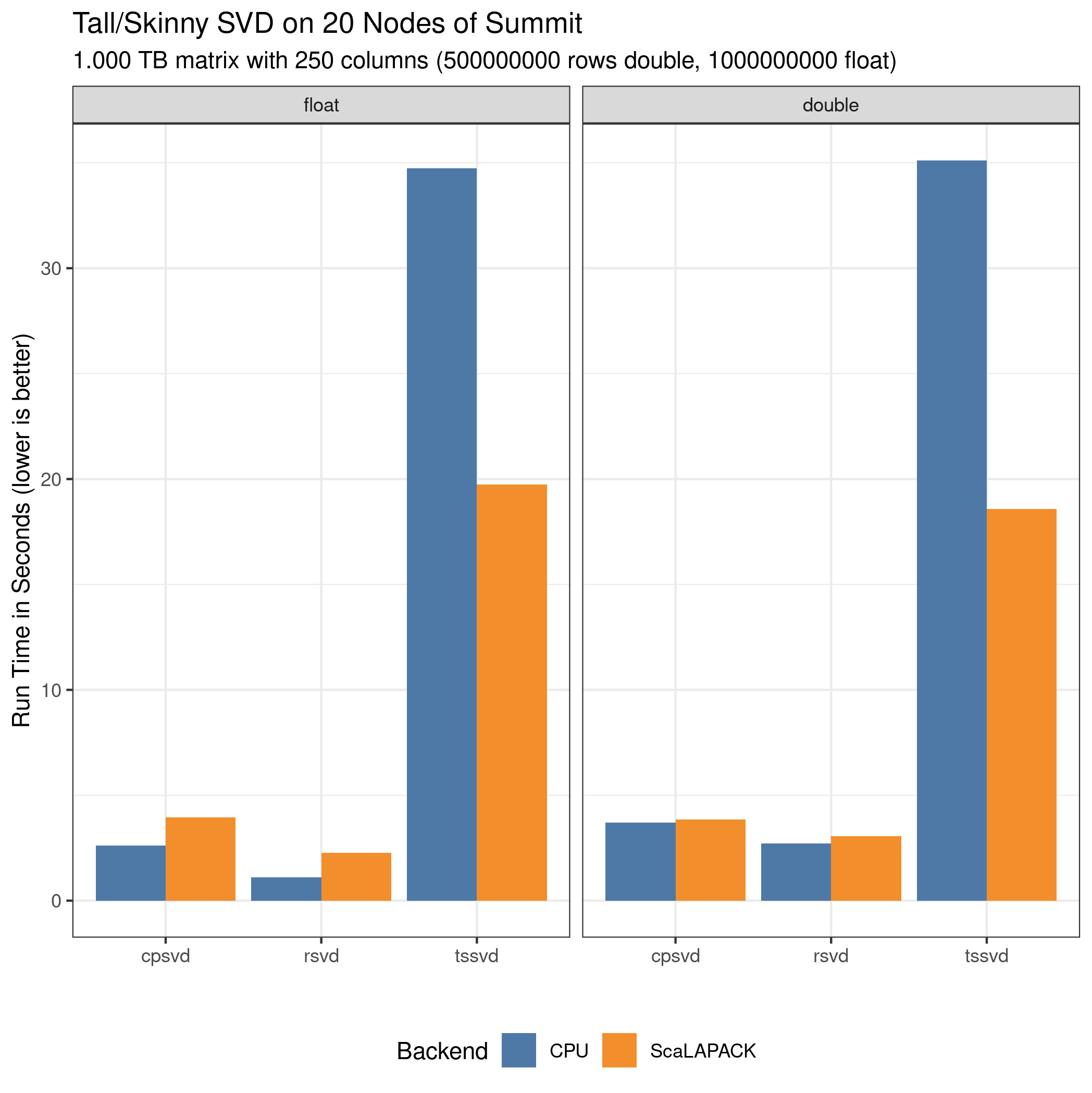}
  \caption{SVD benchmarks compared to ScaLAPACK on 20 Summit nodes. The CPU 
backend refers to our custom implementation, while the ScaLAPACK backend 
refers to an implementation following the same algorithm but with ScaLAPACK 
functions.}
\label{fig:scalapack}
\end{figure}

In our final experiment, we present the results of comparing our approach to 
one using ScaLAPACK functions. In all cases, we use a process grid with one 
column and each matrix has a $16\times 16$ blocking factor. The timings from the 
experiment are shown in Figure~\ref{fig:scalapack}. For the 1 TB problem size 
matrices, the \texttt{pdgesvd} and \texttt{psgesvd} ScaLAPACK functions both 
require an amount of workspace that overflows a 32-bit integer, making it 
impossible to use them. Although this was not done intentionally on our part, it 
is a good demonstration of the need for optimized tall/skinny routines. We 
discuss some possible future strategies for existing ScaLAPACK codebases in 
Section~\ref{sec:conclusions}.

So instead, we use a custom routine which first reduces the matrix using 
\texttt{pdgeqrf} or \texttt{psgeqrf}. Contrary to our tall/skinny SVD using the 
CAQR, these functions use use Householder transformations~\cite{d2000design}. 
For simplicity presenting and discussing the data, we also refer to this 
routine \texttt{tssvd}. For this particular routine, the ScaLAPACK version does 
quite well, being noticeably faster than the alternative. Although this too 
suffers from an indexing issue preventing yet larger experiments, the 
performance of this old library is quite impressive. For the other two 
approaches, our implementation is faster, with the best performance shown in the 
single precision data case. Although in absolute terms, the run times for each 
are small. This shows that so long as matrix sizes conform to those which 
ScaLAPACK is capable of handling and one does not need use of or have access to 
GPUs, using ScaLAPACK is still a very viable choice for many applications nearly 
25 years later.
\section{Conclusions and Future Work}
\label{sec:conclusions}

We have presented a survey of methods for computing the SVD or its 
approximation for distributed tall/skinny data, with particular attention given 
to the detail of implementing these using MPI for communication. We compared 
implementations of these algorithms for both CPU and GPU, single and double 
precision, and presented the results of several experiments from running these 
on the Summit supercomputer.

For local matrix factorizations on GPU, such as computing the 
eigendecomposition in Section~\ref{subsec:normaleqns} or the various QR 
factorizations, we use the vendor library cuSOLVER. Although we have not yet 
conducted serious experiments with it, it would be interesting to use 
MAGMA~\cite{agullo2009numerical} instead. MAGMA provides many advantages over 
cuSOLVER in terms of functionality, and its API is much more similar to LAPACK 
making porting legacy CPU codes to GPUs simpler than using the vendor 
alternative. As far as performance, for some factorizations unrelated to those 
of interest here, the performance of MAGMA seems quite 
good~\cite{haidar2018analysis}; a quick scan of the literature did not produce 
more immediately relevant results. Finally, although cuSOLVER contains only a 
subset of MAGMA functionality, unfortunately swapping cuSOLVER for MAGMA is far 
from a trivial process. 

In Section~\ref{sec:benchmarks}, we compared some of our implementations to 
those using ScaLAPACK functions for the linear algebra operations. ScaLAPACK is 
a very well-written library, but it is showing its age in many ways, notably 
its exclusively 32-bit indexing and inability to utilize GPUs. There is a 
modern replacement for ScaLAPACK called SLATE~\cite{gates2019slate} which 
alleviates these and other issues. Conveniently, SLATE works as an 
\texttt{LD\_PRELOAD} replacement for ScaLAPACK, hijacking symbols at runtime so 
that supported kernels are replaced by high-performance, GPU-enabled variants. 
This would make experimentation for us very simple. However, SLATE is still a
young and developing project. Through private correspondence with one of 
the developers some time ago, the author learned that SLATE does not include 
optimizations for tall/skinny data. Using this without caution could potentially 
result in an experiment which successfully runs and utilizes GPUs, but with 
deceptively poor performance. As SLATE matures, it will be interesting to 
revisit the experiments above.
\section*{Acknowledgements}
We wish to thank George Ostrouchov and Wei-Chen Chen, conversations with whom 
over the past decade have helped shape the author's thinking on this topic.

This research used resources of the Oak Ridge Leadership Computing Facility at 
the Oak Ridge National Laboratory, which is supported by the Office of Science 
of the U.S. Department of Energy under Contract No. DE-AC05-00OR22725.

This manuscript has been authored by UT-Battelle, LLC, under contract 
DE-AC05-00OR22725 with the US Department of Energy (DOE). The US government 
retains and the publisher, by accepting the article for publication, 
acknowledges that the US government retains a nonexclusive, paid-up, 
irrevocable, worldwide license to publish or reproduce the published form of 
this manuscript, or allow others to do so, for US government purposes. DOE will 
provide public access to these results of federally sponsored research in 
accordance with the DOE Public Access Plan 
(\url{http://energy.gov/downloads/doe-public-access-plan}).
%
%
%
%
%
% ------------------------------------------------------------------------
\appendices

\section{Software Environment}
\label{sec:artifacts}

Our implementation is written in C++, and makes extensive use of standard 
interfaces (e.g. BLAS, LAPACK, etc.) and vendor libraries (e.g. cuBLAS, 
cuSOLVER, etc.). All Summit experiments reported above used the following 
software and versions:

\begin{itemize}
    \item gcc 8.1.1
    \item IBM Spectrum MPI 10.3.1.2
    \item NVIDIA CUDA 10.1.243
    \item OpenBLAS 0.3.9
    \item Netlib ScaLAPACK 2.0.2
    \item \texttt{CXXFLAGS = -O2 -std=c++17 -mcpu=native =mtune=native}
    \item OpenMP (for \texttt{omp simd})
    \item \texttt{OMP\_NUM\_THREADS=1}
\end{itemize}
%
%
%
%
%
% ------------------------------------------------------------------------
\bibliographystyle{IEEEtran}
\bibliography{svd}

\begin{thebibliography}{10}
\providecommand{\url}[1]{#1}
\csname url@samestyle\endcsname
\providecommand{\newblock}{\relax}
\providecommand{\bibinfo}[2]{#2}
\providecommand{\BIBentrySTDinterwordspacing}{\spaceskip=0pt\relax}
\providecommand{\BIBentryALTinterwordstretchfactor}{4}
\providecommand{\BIBentryALTinterwordspacing}{\spaceskip=\fontdimen2\font plus
\BIBentryALTinterwordstretchfactor\fontdimen3\font minus
  \fontdimen4\font\relax}
\providecommand{\BIBforeignlanguage}[2]{{%
\expandafter\ifx\csname l@#1\endcsname\relax
\typeout{** WARNING: IEEEtran.bst: No hyphenation pattern has been}%
\typeout{** loaded for the language `#1'. Using the pattern for}%
\typeout{** the default language instead.}%
\else
\language=\csname l@#1\endcsname
\fi
#2}}
\providecommand{\BIBdecl}{\relax}
\BIBdecl

\bibitem{mccullagh1989generalized}
P.~McCullagh and J.~A. Nelder, ``Generalized linear models 2nd edition chapman
  and hall,'' \emph{London, UK}, 1989.

\bibitem{wall2003singular}
M.~E. Wall, A.~Rechtsteiner, and L.~M. Rocha, ``Singular value decomposition
  and principal component analysis,'' in \emph{A practical approach to
  microarray data analysis}.\hskip 1em plus 0.5em minus 0.4em\relax Springer,
  2003, pp. 91--109.

\bibitem{MPI1994}
W.~Gropp, E.~Lusk, and A.~Skjellum, \emph{Using {MPI}: Portable Parallel
  Programming with the Message-Passing Interface}.\hskip 1em plus 0.5em minus
  0.4em\relax Cambridge, MA, USA: MIT Press Scientific And Engineering
  Computation Series, 1994.

\bibitem{dean2010mapreduce}
J.~Dean and S.~Ghemawat, ``Mapreduce: a flexible data processing tool,''
  \emph{Communications of the ACM}, vol.~53, no.~1, pp. 72--77, 2010.

\bibitem{constantine2011tall}
P.~G. Constantine and D.~F. Gleich, ``Tall and skinny qr factorizations in
  mapreduce architectures,'' in \emph{Proceedings of the second international
  workshop on MapReduce and its applications}, 2011, pp. 43--50.

\bibitem{darema2001spmd}
F.~Darema, ``The spmd model: Past, present and future,'' in \emph{European
  Parallel Virtual Machine/Message Passing Interface Users’ Group
  Meeting}.\hskip 1em plus 0.5em minus 0.4em\relax Springer, 2001, pp. 1--1.

\bibitem{tu2008scalable}
T.~Tu, C.~A. Rendleman, D.~W. Borhani, R.~O. Dror, J.~Gullingsrud, M.~O.
  Jensen, J.~L. Klepeis, P.~Maragakis, P.~Miller, K.~A. Stafford \emph{et~al.},
  ``A scalable parallel framework for analyzing terascale molecular dynamics
  simulation trajectories,'' in \emph{SC'08: Proceedings of the 2008 ACM/IEEE
  conference on Supercomputing}.\hskip 1em plus 0.5em minus 0.4em\relax IEEE,
  2008, pp. 1--12.

\bibitem{plimpton2011mapreduce}
S.~J. Plimpton and K.~D. Devine, ``Mapreduce in mpi for large-scale graph
  algorithms,'' \emph{Parallel Computing}, vol.~37, no.~9, pp. 610--632, 2011.

\bibitem{zaharia2016apache}
M.~Zaharia, R.~S. Xin, P.~Wendell, T.~Das, M.~Armbrust, A.~Dave, X.~Meng,
  J.~Rosen, S.~Venkataraman, M.~J. Franklin \emph{et~al.}, ``Apache spark: a
  unified engine for big data processing,'' \emph{Communications of the ACM},
  vol.~59, no.~11, pp. 56--65, 2016.

\bibitem{xenopoulos2016big}
P.~Xenopoulos, J.~Daniel, M.~Matheson, and S.~Sukumar, ``Big data analytics on
  hpc architectures: Performance and cost,'' in \emph{2016 IEEE International
  Conference on Big Data (Big Data)}.\hskip 1em plus 0.5em minus 0.4em\relax
  IEEE, 2016, pp. 2286--2295.

\bibitem{booth2006power}
T.~E. Booth, ``Power iteration method for the several largest eigenvalues and
  eigenfunctions,'' \emph{Nuclear science and engineering}, vol. 154, no.~1,
  pp. 48--62, 2006.

\bibitem{lanczos1950iteration}
C.~Lanczos, \emph{An iteration method for the solution of the eigenvalue
  problem of linear differential and integral operators}.\hskip 1em plus 0.5em
  minus 0.4em\relax United States Governm. Press Office Los Angeles, CA, 1950.

\bibitem{demmel1997applied}
J.~W. Demmel, \emph{Applied numerical linear algebra}.\hskip 1em plus 0.5em
  minus 0.4em\relax SIAM, 1997.

\bibitem{berry1995using}
M.~W. Berry, S.~T. Dumais, and G.~W. O’Brien, ``Using linear algebra for
  intelligent information retrieval,'' \emph{SIAM review}, vol.~37, no.~4, pp.
  573--595, 1995.

\bibitem{page1999pagerank}
L.~Page, S.~Brin, R.~Motwani, and T.~Winograd, ``The pagerank citation ranking:
  Bringing order to the web.'' Stanford InfoLab, Tech. Rep., 1999.

\bibitem{halko2011finding}
N.~Halko, P.-G. Martinsson, and J.~A. Tropp, ``Finding structure with
  randomness: Probabilistic algorithms for constructing approximate matrix
  decompositions,'' \emph{SIAM review}, vol.~53, no.~2, pp. 217--288, 2011.

\bibitem{lawson1979basic}
C.~L. Lawson, R.~J. Hanson, D.~R. Kincaid, and F.~T. Krogh, ``Basic linear
  algebra subprograms for fortran usage,'' \emph{ACM Transactions on
  Mathematical Software (TOMS)}, vol.~5, no.~3, pp. 308--323, 1979.

\bibitem{lug}
E.~Anderson, Z.~Bai, C.~Bischof, L.~S. Blackford, J.~Dem~mel, J.~J. Dongarra,
  J.~Du~Croz, S.~Hammarling, A.~Greenbaum, A.~McKenney, and D.~Sorensen,
  \emph{LAPACK Users' guide (third ed.)}.\hskip 1em plus 0.5em minus
  0.4em\relax Philadelphia, PA, USA: Society for Industrial and Applied
  Mathematics, 1999.

\bibitem{xianyi2012openblas}
Z.~Xianyi, W.~Qian, and Z.~Chothia, ``Openblas,'' \emph{URL: http://xianyi.
  github. io/OpenBLAS}, p.~88, 2012.

\bibitem{kirk2007nvidia}
D.~Kirk \emph{et~al.}, ``Nvidia cuda software and gpu parallel computing
  architecture,'' in \emph{ISMM}, vol.~7, 2007, pp. 103--104.

\bibitem{team2000r}
R.~C. Team, ``R language definition,'' \emph{Vienna, Austria: R foundation for
  statistical computing}, 2000.

\bibitem{golub1996matrix}
G.~Golub and C.~Van~Loan, ``Matrix computations 3rd edition the john hopkins
  university press,'' \emph{Baltimore, MD}, 1996.

\bibitem{demmel2008communication}
J.~Demmel, L.~Grigori, M.~Hoemmen, and J.~Langou, ``Communication-avoiding
  parallel and sequential qr factorizations,'' \emph{CoRR abs/0806.2159}, 2008.

\bibitem{agullo2010qr}
E.~Agullo, C.~Coti, J.~Dongarra, T.~Herault, and J.~Langem, ``Qr factorization
  of tall and skinny matrices in a grid computing environment,'' in \emph{2010
  IEEE International Symposium on Parallel \& Distributed Processing
  (IPDPS)}.\hskip 1em plus 0.5em minus 0.4em\relax IEEE, 2010, pp. 1--11.

\bibitem{rabenseifner2004optimization}
R.~Rabenseifner, ``Optimization of collective reduction operations,'' in
  \emph{International Conference on Computational Science}.\hskip 1em plus
  0.5em minus 0.4em\relax Springer, 2004, pp. 1--9.

\bibitem{schmidt2017programming}
D.~Schmidt, W.-C. Chen, M.~A. Matheson, and G.~Ostrouchov, ``Programming with
  big data in r: Scaling analytics from one to thousands of nodes,'' \emph{Big
  Data Research}, vol.~8, pp. 1--11, 2017.

\bibitem{slug}
\BIBentryALTinterwordspacing
L.~S. Blackford, J.~Choi, A.~Cleary, E.~D'Azevedo, J.~Demmel, I.~Dhillon,
  J.~Dongarra, S.~Hammarling, G.~Henry, A.~Petitet, K.~Stanley, D.~Walker, and
  R.~C. Whaley, \emph{{ScaLAPACK} Users' Guide}.\hskip 1em plus 0.5em minus
  0.4em\relax Philadelphia, PA: Society for Industrial and Applied Mathematics,
  1997. [Online]. Available:
  \url{http://netlib.org/scalapack/slug/scalapack_slug.html/}
\BIBentrySTDinterwordspacing

\bibitem{dongarra1997top500}
J.~J. Dongarra, H.~W. Meuer, E.~Strohmaier \emph{et~al.}, ``Top500
  supercomputer sites,'' \emph{Supercomputer}, vol.~13, pp. 89--111, 1997.

\bibitem{top500june2020}
``{TOP500} june 2020,'' \url{https://top500.org/lists/top500/2020/06/},
  accessed: 2020-08-18.

\bibitem{nsight20133}
N.~Nsight and V.~S. Edition, ``3.0 user guide,'' \emph{NVIDIA Corporation},
  2013.

\bibitem{d2000design}
E.~D'Azevedo and J.~Dongarra, ``The design and implementation of the parallel
  out-of-core scalapack lu, qr, and cholesky factorization routines,''
  \emph{Concurrency: Practice and Experience}, vol.~12, no.~15, pp. 1481--1493,
  2000.

\bibitem{agullo2009numerical}
E.~Agullo, J.~Demmel, J.~Dongarra, B.~Hadri, J.~Kurzak, J.~Langou, H.~Ltaief,
  P.~Luszczek, and S.~Tomov, ``Numerical linear algebra on emerging
  architectures: The plasma and magma projects,'' in \emph{Journal of Physics:
  Conference Series}, vol. 180, no.~1.\hskip 1em plus 0.5em minus 0.4em\relax
  IOP Publishing, 2009, p. 012037.

\bibitem{haidar2018analysis}
A.~Haidar, S.~Tomov, J.~Dongarra \emph{et~al.}, ``Analysis and design
  techniques towards high-performance and energy-efficient dense linear solvers
  on gpus,'' \emph{IEEE Transactions on Parallel and Distributed Systems},
  vol.~29, no.~12, pp. 2700--2712, 2018.

\bibitem{gates2019slate}
M.~Gates, J.~Kurzak, A.~Charara, A.~YarKhan, and J.~Dongarra, ``Slate: design
  of a modern distributed and accelerated linear algebra library,'' in
  \emph{Proceedings of the International Conference for High Performance
  Computing, Networking, Storage and Analysis}, 2019, pp. 1--18.

\end{thebibliography}

\end{document}